\documentclass[aps,prapplied,reprint,superscriptaddress]{revtex4-2}
\usepackage{graphicx}
\usepackage{physics}
\usepackage[hidelinks]{hyperref}

\usepackage{titlesec}
\titlespacing*{\section}{0pt}{3ex plus 0.5ex minus 0.2ex}{0.2ex}
\titlespacing*{\subsection}{0pt}{3ex plus 0.5ex minus 0.2ex}{0.2ex}

\begin{document}

\title[Complex-amplitude Fourier single-pixel imaging]{Complex-amplitude
  Fourier single-pixel imaging via coherent structured illumination}

\author{Ya-Nan Zhao}
\affiliation{Hebei Key Laboratory of Optic-Electronic Information and
  Materials, College of Physics Science \& Technology, Hebei University,
  Baoding 071002, China}

\author{Hong-Yun Hou}
\affiliation{Hebei Key Laboratory of Optic-Electronic Information and
  Materials, College of Physics Science \& Technology, Hebei University,
  Baoding 071002, China}

\author{Jia-Cheng Han}
\affiliation{Hebei Key Laboratory of Optic-Electronic Information and
  Materials, College of Physics Science \& Technology, Hebei University,
  Baoding 071002, China}

\author{De-Zhong Cao}
\affiliation{Department of Physics, Yantai University, Yantai 264005, China}

\author{Su-Heng Zhang}
\email{shzhang@hbu.edu.cn}
\affiliation{Hebei Key Laboratory of Optic-Electronic Information and
  Materials, College of Physics Science \& Technology, Hebei University,
  Baoding 071002, China}

\author{Hong-Chao Liu}
\email{hcliu@um.edu.mo}
\affiliation{Institute of Applied Physics and Materials Engineering,
  University of Macau, Avenida da Universidade, Taipa, Macao SAR, China}

\author{Bao-Lai Liang}
\email{liangbaolai@gmail.com}
\affiliation{Hebei Key Laboratory of Optic-Electronic Information and
  Materials, College of Physics Science \& Technology, Hebei University,
  Baoding 071002, China}

\date{\today}

\begin{abstract}
  We propose a method of complex-amplitude Fourier single-pixel imaging (CFSI)
  with coherent structured illumination to acquire both the amplitude and phase
  of an object.
  In the proposed method, an object is illustrated by a series of coherent
  structured light fields which are generated by a phase-only spatial light
  modulator, the complex Fourier spectrum of the object can be acquired
  sequentially by a single-pixel photodetector.
  Then the desired complex-amplitude image can be retrieved directly by
  applying an inverse Fourier transform.
  We experimentally implemented this CFSI with several different types
  of objects.
  The experimental results show that the proposed method provides a promising
  complex-amplitude imaging approach with high quality and a stable
  configuration.
  Thus, it might find broad applications in optical metrology and
  biomedical science.
\end{abstract}

\maketitle

\section{INTRODUCTION}
An optical field is expressed as a complex-amplitude, which describes both
the amplitude and phase of the light wave.
Conventional imaging only observes amplitude information,
but the important phase information is lost.
This is because from the photodetector to the human retina only respond to
light intensity.
Developing an efficient approach to recover full complex-amplitude of
an optical field has been one of the most attractive challenges
in modern imaging science.
Starting with early Zernike's phase contrast microscopy
\cite{zernikePhaseContrastNew1942},
various complex-amplitude imaging techniques have been proposed, such as
differential interference contrast microscopy
\cite{nomarski1955differential},
Shack-Hartmann sensing
\cite{shack1971production},
coherent diffraction imaging
\cite{miaoExtendingMethodologyXray1999},
digital holography
\cite{marquetDigitalHolographicMicroscopy2005},
Fourier ptychographic microscopy
\cite{zhengWidefieldHighresolutionFourier2013},
lensless ghost imaging
\cite{zhangLenslessGhostImaging2014,
  yuFourierTransformGhostImaging2016,
  zhaoCorrelatedReconstructionPhaseonly2021},
and phase imaging techniques based on
transport-of-intensity equation
\cite{zuoTransportIntensityEquation2020}.

However, almost all the above methods require pixelated imaging sensors.
This leads to a strong challenge for the cases with a light of invisible
wavelength or extremely low intensity.
Because it can be impractical or prohibitively costly to implement with
a pixelated imaging device.
Single-pixel imaging (SPI), as an emerging imaging technique characterized by
using structured illumination and a single-pixel detector,
has the potential to overcome the challenge
\cite{gibsonSinglepixelImaging122020}.
Thus, several attempts were made to achieve the complex amplitude information
of objects by using SPI techniques, such as
single-pixel diffractive imaging
\cite{leeSinglepixelCoherentDiffraction2010,
  horisakiSinglepixelCompressiveDiffractive2017a,
  horisakiSinglepixelCompressiveDiffractive2017},
single-pixel wavefront sensing
\cite{howlandCompressiveWavefrontSensing2014,
  shinReferenceFreeSinglePointHolographic2018,
  soldevilaPhaseImagingSpatial2018},
single-pixel ptychography
\cite{liSinglepixelPtychography2021},
single-pixel digital holography
\cite{clementeCompressiveHolographySinglepixel2013,
  martinez-leonSinglepixelDigitalHolography2017,
  gonzalezHighSamplingRate2018,
  huSinglepixelPhaseImaging2019,
  santos-amadorPhaseAmplitudeReconstruction2021,
  wuImagingBiologicalTissue2021},
and single-pixel phase imaging based on common-path interferometry
\cite{liuSinglepixelPhaseFluorescence2018,
  liuComplexWavefrontReconstruction2019,
  zhaoFourierSinglepixelReconstruction2019,
  liuSinglepixelSpiralPhase2020,
  liQuantitativeImagingOptical2021,
  houComplexamplitudeSinglepixelImaging2021}.
Benefiting from the advantages of single-pixel detectors, the complex-amplitude
single-pixel imaging is becoming a promising imaging modality
in the fields of optical microscopy
\cite{liuSinglepixelPhaseFluorescence2018},
optical metrology
\cite{liQuantitativeImagingOptical2021},
and biomedical science
\cite{wuImagingBiologicalTissue2021}.

In this letter, we present a novel complex-amplitude Fourier single-pixel
imaging (CFSI) method by combining coherent structured illumination
\cite{martinez-leonSinglepixelDigitalHolography2017,
  horisakiSinglepixelCompressiveDiffractive2017,
  liuSinglepixelPhaseFluorescence2018,
  zhaoFourierSinglepixelReconstruction2019,
  liQuantitativeImagingOptical2021}
and common-path interference
\cite{liuSinglepixelPhaseFluorescence2018,
  liuComplexWavefrontReconstruction2019,
  zhaoFourierSinglepixelReconstruction2019,
  liuSinglepixelSpiralPhase2020,
  liQuantitativeImagingOptical2021,
  houComplexamplitudeSinglepixelImaging2021}
with Fourier basis scan
\cite{zhangSinglepixelImagingMeans2015,
  huangComputationalweightedFourierSinglepixel2018,
  zhangSimultaneousSpatialSpectral2018,
  pengFourierMicroscopyBased2021}.
Different from previous CFSI methods
\cite{huSinglepixelPhaseImaging2019,
  zhaoFourierSinglepixelReconstruction2019,
  liuSinglepixelSpiralPhase2020}
that use the digital micromirror device based on the super-pixel method
for complex-amplitude modulation,
we employ a phase-only spatial light modulator to generate both
the structured light and the reference light to form
the coherent structured illumination.
The phase modulation proposal is relatively simple and efficient.
Compared to a two-beam interferometer, the single-beam structure is more
compact and stable in practical application.
With the help of a 4-step phase-shifting, the complex-valued Fourier
spectrum can be directly acquired by single-pixel detection.
The desired complex-amplitude image can be further retrieved by applying an
inverse Fourier transform.
In the experiments, the proposed CFSI is implemented with three different
types of objects.
Experimental results demonstrate that the coherent structured illumination
enable the CFSI method to have high quality and a stable configuration.
In addition, we find that the undersampling technique can effectively
remove noise and considerably accelerate the image acquisition process
in our CFSI scheme.

\section{METHODS}
\subsection{Schematic of our CFSI}
\begin{figure}[tbp]
\centering
\includegraphics[width=\columnwidth]{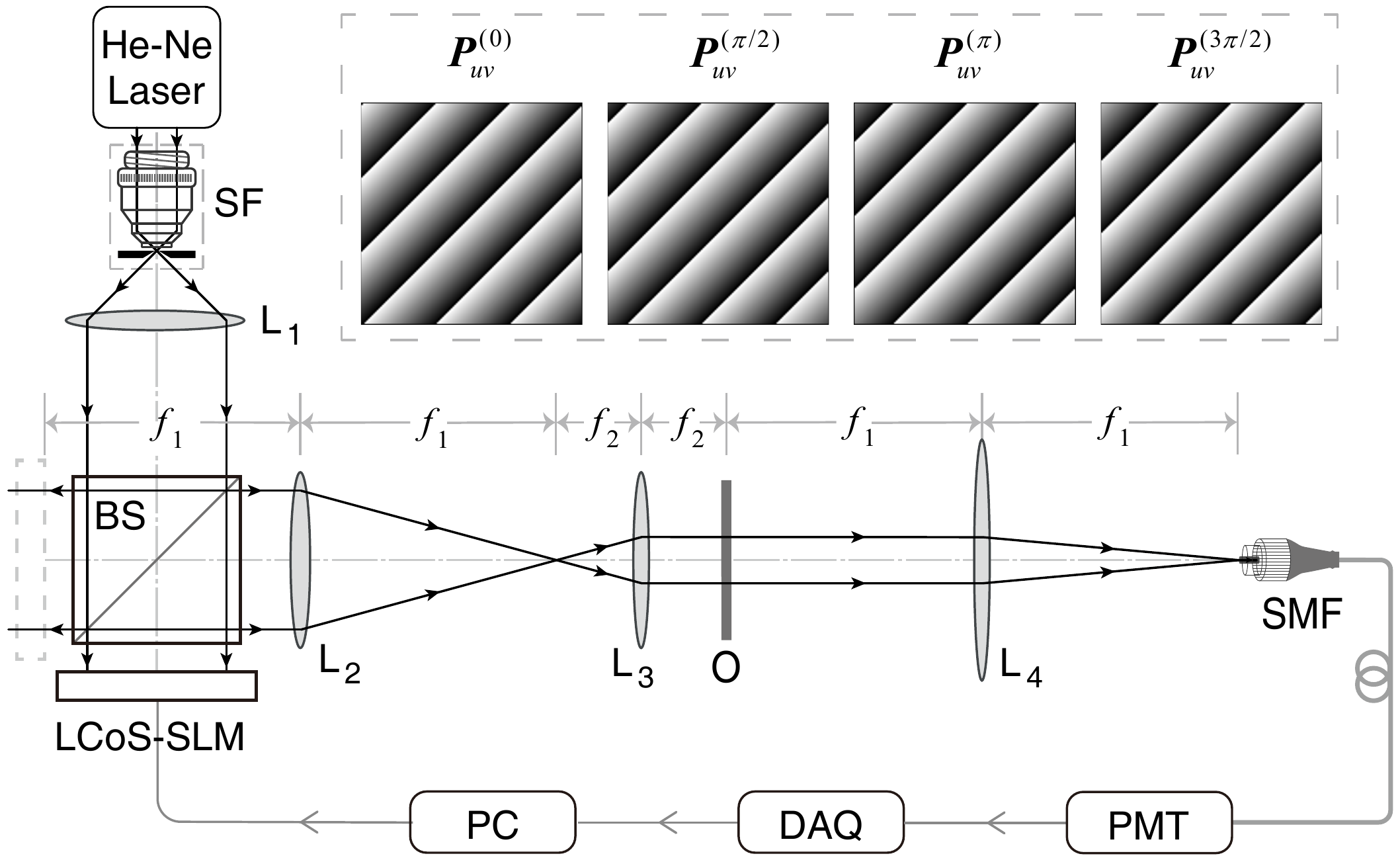}
\caption{\label{fig_scheme}Schematic of our CFSI.
  SF, spatial filter system;
  L$_1$, collimating lens;
  BS, 50:50 non-polarizing beamsplitter;
  LCoS-SLM, liquid-crystal-on-silicon spatial light modulator;
  L$_2$, L$_3$, achromatic doublet lenses;
  O, target object;
  L$_4$, collecting lens;
  SMF, single mode fiber optic patch cable;
  PMT, photomultiplier tube;
  DAQ, data acquisition board;
  PC, personal computer.}
\end{figure}

The schematic of our CFSI is depicted in Fig. \ref{fig_scheme}.
A He-Ne laser of wavelength $ 632.8 $ nm is first expanded and
collimated by a spatial filter (SF) and a collimating lens (L$_1$),
then passes through a beam splitter (BS),
and finally incident on a phase-only liquid-crystal-on-silicon
spatial light modulator (LCoS-SLM).
Since the modulation efficiency of the LCoS is not $100$\%, the reflected
light consists of two parts, the phase-modulated structured light
and the directly reflected light that serves as the reference light.
The structured light and the reference light travel along the same path,
resulting in common-path interference to form coherent structured illumination.
The coherent structured illumination is reflected by BS and imaged onto
the target object (O) via a $ 4f $ system, which consists of two achromatic
doublet lenses (L$_2$, L$_3$) with an aperture of $ 50 $ mm and focal lengths
of $ 30 $ cm and $ 10 $ cm, respectively.
The transmitted object light passes through a collecting lens (L$_4$)
of focal length $ 30 $ cm and aperture $ 75 $ mm,
and then reaches the single mode fiber optic patch cable (SMF).
Since O and the end face of SMF are located at the front and rear focal
planes of L$_4$, the Fourier spectrum of the transmitted object light
is obtained at the end face of SMF.
Due to the core diameter of about $ 9 $ {\textmu}m, SMF collects only
the zero-frequency light.
A photomultiplier tube (PMT) detects the light from SMF,
and the resulting output is fed to PC via a data acquisition board (DAQ).

\subsection{Fourier basis scan for the CFSI}
Let the spatial sampling matrix of the complex transmittance of the target
object be $ \vb*{f} $.
By definition, the two-dimensional discrete Fourier transform can be
written as
\begin{equation}
  F(u,v)=\sum^{M-1}_{x=0}\sum^{N-1}_{y=0}f(x,y)
  \exp[ -j2\pi\left(\frac{ux}{M}+\frac{vy}{N}\right) ],
\end{equation}
where $ f(x,y) $ represents the element of the sampling matrix $ \vb*{f} $,
$ F(u,v) $ represents the Fourier spectrum component,
and $ u=0,1,\cdots,M-1 $, $ v=0,1,\cdots,N-1 $, and $ M ,\; N $ denote
the number of rows and columns of the sampling matrix $ \vb*{f} $,
respectively.
The complete set of Fourier basis matrices can be expressed as
\begin{equation}
\left\{
  \vb*{T}_{uv}\Bigg|\,
  \mqty{u=0,1,\cdots,\left\lceil{M/2}\right\rceil-1,
      -\left\lfloor{M/2}\right\rfloor,\cdots,-1,\\
        v=0,1,\cdots,\left\lceil{N/2}\right\rceil-1,\,
      -\left\lfloor{N/2}\right\rfloor,\,\cdots,-1.}
\right\},
\end{equation}
where the Fourier basis matrix $ \vb*{T}_{uv} $ is defined as
\begin{equation}\label{eq_FBasisDef}
  T_{uv}(x,y)=
  \exp[-j2\pi\left(\frac{ux}{M}+\frac{vy}{N}\right)],
\end{equation}
where
\begin{align}
  x&=-\left\lfloor{M/2}\right\rfloor,\cdots,-1,0,1,\cdots,
  \left\lceil{M/2}\right\rceil-1,\notag\\
  y&=-\left\lfloor{N/2}\right\rfloor,\cdots,-1,0,1,\cdots,
  \left\lceil{N/2}\right\rceil-1,\notag
\end{align}
where $ \left\lfloor\cdot\right\rfloor $ and $ \left\lceil\cdot\right\rceil $
denote floor and ceiling operations, respectively.
Taking into account the periodicity of the discrete Fourier transform,
the range of values of the spatial coordinates is shifted so that
the origin of the coordinates lies at the center of the matrix,
and the range of values of the spatial frequency coordinates is also shifted
to facilitate the inverse discrete Fourier transform (IDFT) to reconstruct the
complex transmittance of the target object.
Then the Fourier spectrum coefficient $ F(u,v) $ can be obtained by
\begin{equation}\label{eq_FBasiScan}
  F(u,v) =\langle\vb*{f}, \vb*{T}^*_{uv}\rangle_{\mathrm{F}},
\end{equation}
where $ \langle\cdot,\cdot\rangle_\mathrm{F}$ denotes the Frobenius inner
product, and $ \ast $ denotes complex conjugate.
Each Fourier spectrum coefficient can be acquired one by one, hence this
technique is called Fourier basis scan.

From the definition of Eq. (\ref{eq_FBasisDef}), it can be seen that the
Fourier basis matrix $ \vb*{T}_{uv} $ is a phase-only matrix with the phase
distribution of $ \mathrm{arg}\vb*{T}_{uv} \in [0,2\pi) $.
Using the phase distribution, we can generate a phase pattern
\begin{equation}
  \vb*{P}_{uv}= \frac{1}{2\pi}\mathrm{arg}\vb*{T}_{uv}.
\end{equation}
We load the LCoS with the phase pattern $ \vb*{P}_{uv} $ to generate the
corresponding structured light,
which has the form of $ E_0\vb*{T}_{uv} $, where $ E_0 $ is the amplitude of
the structured light.
The reference light can be expressed as $ E_r e^{-j\phi_r} $, where $ E_r $ is
the amplitude and $ \phi_r $ is the initial phase.
Thus, the coherent structured illumination has the form of
\begin{equation}
\vb*{E}_{uv}=E_0\vb*{T}_{uv}+E_r e^{-j\phi_r}\vb*{1},
\end{equation}
where $ \vb*{1} $ denotes a matrix of ones which has the same dimension as
$ \vb*{T}_{uv} $.
Under this coherent structured illumination, the zero-frequency component
of the object light measured by PMT can be expressed as
\begin{equation}
  D(u,v)=\eta\left|\left\langle \vb*{f}, \vb*{E}^*_{uv}
    \right\rangle_{\mathrm{F}}\right|,
\end{equation}
where $ \eta $ represents the quantum efficiency of PMT.

A 4-step phase shift method is used to acquire the Fourier spectrum.
This requests shifting the Fourier basis matrix $ \vb*{T}_{uv} $ by
$ \varphi= 0 $, $ \pi/2 $, $ \pi $, $ 3\pi/2 $ to obtain a set of phase-shifted
basis matrices
$ \vb*{T}^{(\varphi)}_{uv} = e^{j\varphi}\vb*{T}_{uv}$.
The corresponding phase patterns are generated by
$ \vb*{P}^{(\varphi)}_{uv} = \mathrm{arg}\vb*{T}^{(\varphi)}_{uv}/2\pi $,
as shown in the inset of Fig. \ref{fig_scheme}.
We load the whole set of the phase patterns
$ \big\{ \vb*{P}^{(\varphi)}_{uv} \big\} $
onto the LCoS one by one and acquire the corresponding responses of
PMT, denoted as $ \big\{ \vb*{D}^{(\varphi)}(u,v)\big\} $.
Then we can retrieve the Fourier spectrum matrix of the target object
from the responses of PMT as
\begin{equation}
  \vb*{F}=\alpha \left\{\big[\vb*{D}^{(0)}-\vb*{D}^{(\pi)}\big]
  +j\big[\vb*{D}^{(3\pi/2)}-\vb*{D}^{(\pi/2)}\big]\right\},
\end{equation}
where $ \alpha $ is a complex constant.
By using the IDFT, we can reconstruct the complex transmittance of the
target object
\begin{equation}
  \vb*{f}=\mathrm{fftshift}\!\big\{\mathrm{ifft2}\!\{\vb*{F}\}\big\},
\end{equation}
where $ \mathrm{ifft2}\!\{\cdot\} $ denotes the two-dimensional
IDFT operation using a fast algorithm
and the subsequent $\mathrm{fftshift}\!\{\cdot\}$ operation shifts
the coordinate origin to the center of the matrix.

\section{RESULTS}
In all experiments, we set the imaging resolution to
$128 \times 128$ pixels, and display each pixel of the phase patterns
as $ 2 \times 2 $ LCoS pixels.
Since the effective aperture is large enough, all the coherent structured
illumination patterns can be imaged onto the object by the $4f$ system.
As the pixel pitch of the LCoS is $ 8 $ {\textmu}m and the magnification of
the $4f$ system is $1/3$, the spatial resolution and the field of view
of this CFSI are theoretically $ 5.33 $ {\textmu}m and $ 682.67 $ {\textmu}m
in the experiments.

\begin{figure}[htbp]
\centering
\includegraphics[width=\columnwidth]{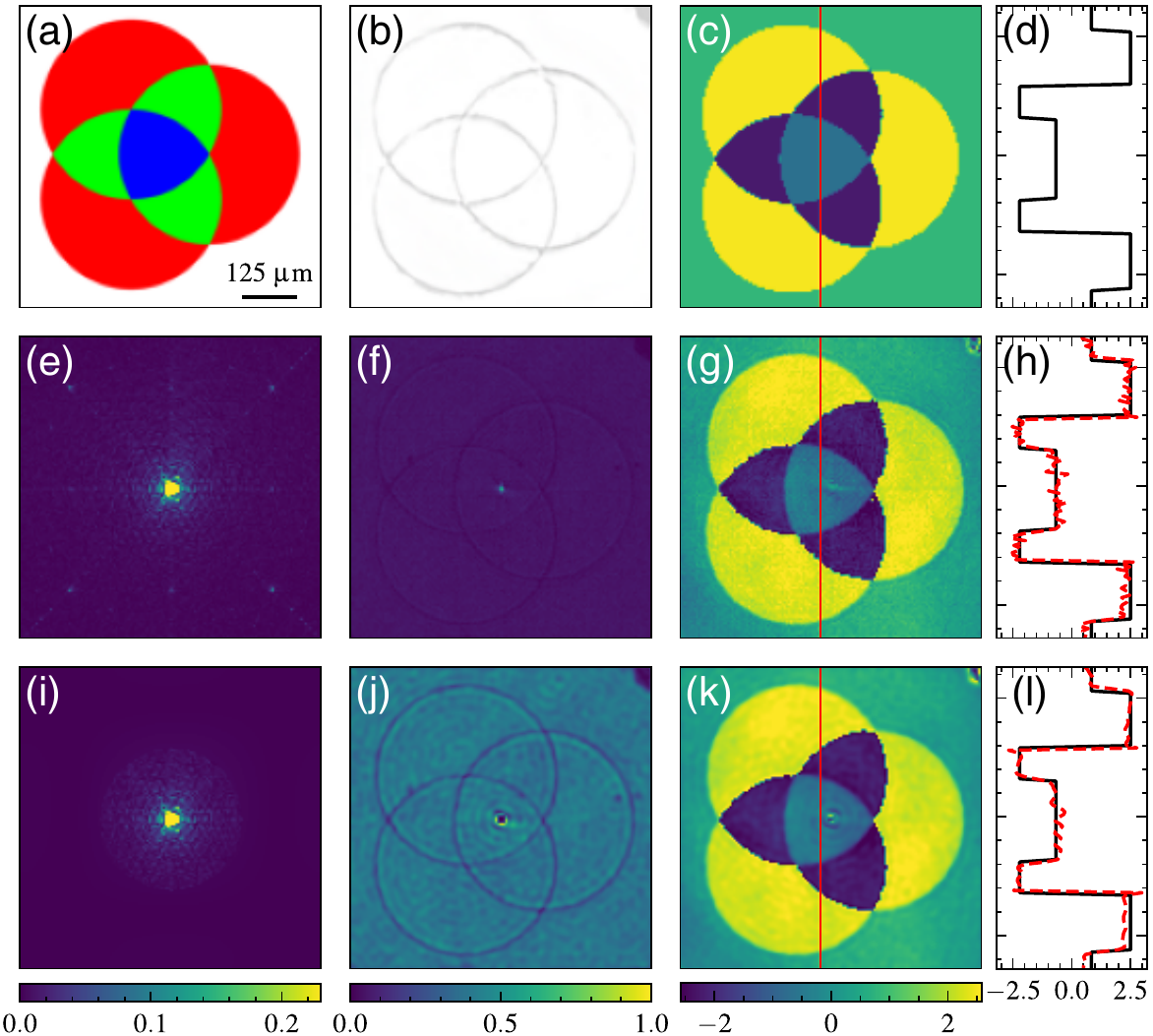}
\caption{\label{fig_disc}The glass plate etched with three intersecting discs
  and related experimental results.
  (a) Geometry diagram of the etched discs.
  (b) The macro photo of the etched object.
  (c) The theoretical phase distribution of the etched object.
  (d) The phase profile along the highlighted line shown in (c).
  (e)--(h) are the magnitude of the measured spectrum, the reconstructed
  amplitude and phase images, and the phase profile at full sampling ratio,
  respectively, and
  (i)--(l) are the corresponding results at the sampling ratio of $ 17.2 $\%,
  respectively. See Data File 1 for underlying values.}
\end{figure}

As a demonstration, we first apply our CFSI to image a glass plate
etched with three intersecting discs.
As shown in Fig. \ref{fig_disc}(a), the discs are all $400$ {\textmu}m
in diameter, and the circumference of each disc passes through the centers
of the other two discs.
The etching depths in red, green and blue regions are $372$ nm, $715$ nm,
and $1051$ nm, respectively.
This is a simple phase object.
Fig. \ref{fig_disc}(b) shows the macro photo of this etched object
captured by a digital camera.
Fig. \ref{fig_disc}(c) shows the theoretical phase distribution of this
etched object for the light of wavelength $632.8$ nm,
and the phase profile along the highlighted line is shown in
Fig. \ref{fig_disc}(d).

Fig. \ref{fig_disc}(e) shows the magnitude of the measured spectrum.
The values of the measured spectrums are linearly scaled to $ [0,\,1] $,
but the dynamic range of the spectrum is compressed to show
the high frequency components more clearly.
The reconstructed amplitude and phase images are shown in
Figs. \ref{fig_disc} (f) and (g), respectively,
and the phase profile along the highlighted line is shown in
Fig. \ref{fig_disc}(h).
The values of the reconstructed amplitude images are linearly
scaled to $ [0,\,1] $ for display.
Due to the phase modulation error of LCoS-SLM, external vibration and
other factors, the measured spectrum contains noises, which affect
the reconstructed image quality.
The low frequency noises cause a bright dot in the center of
the reconstructed amplitude image,
as shown in Fig. \ref{fig_disc}(f).
The high frequency noises, which can be seen as symmetric pairs of
bright dots in the magnitude spectrum shown in Fig. \ref{fig_disc}(e),
cause the reconstructed phase fluctuation,
as shown in Fig. \ref{fig_disc}(h).
At full sampling ratio, the imaging acquisition time is $ 195 $ minutes,
which is mainly limited by the response time and phase flicker of LCoS-SLM.

The high frequency noises can be avoided by undersampling.
Here we set the sampling ratio of $17.2$\%, at which point the high frequency
noises basically disappears, as shown in Fig. \ref{fig_disc}(i).
In addition, undersampling can significantly reduce the acquisition time.
At the current sampling ratio, the imaging acquisition time is
about $34$ minutes.
To eliminate the influence of low frequency noises, we need perform
noise suppression during image reconstruction,
similar to that proposed by
Xiao \emph{et al.} in reference\cite{xiaoDirectSingleStepMeasurement2019}.
We use the subimage within the $3\times3$ neighborhood of the center of
the reconstructed image as an estimate of the spectrum of
the low frequency noises in all experiments.
The final reconstructed amplitude and phase images are shown in
Figs. \ref{fig_disc}(j) and (k), respectively,
and the phase profile along the highlighted line is shown in
Fig. \ref{fig_disc}(l).
It can be seen that the reconstructed image after noise reduction has clear
amplitude and smooth and accurate phase.

\begin{figure}[tbp]
\centering
\includegraphics[width=\columnwidth]{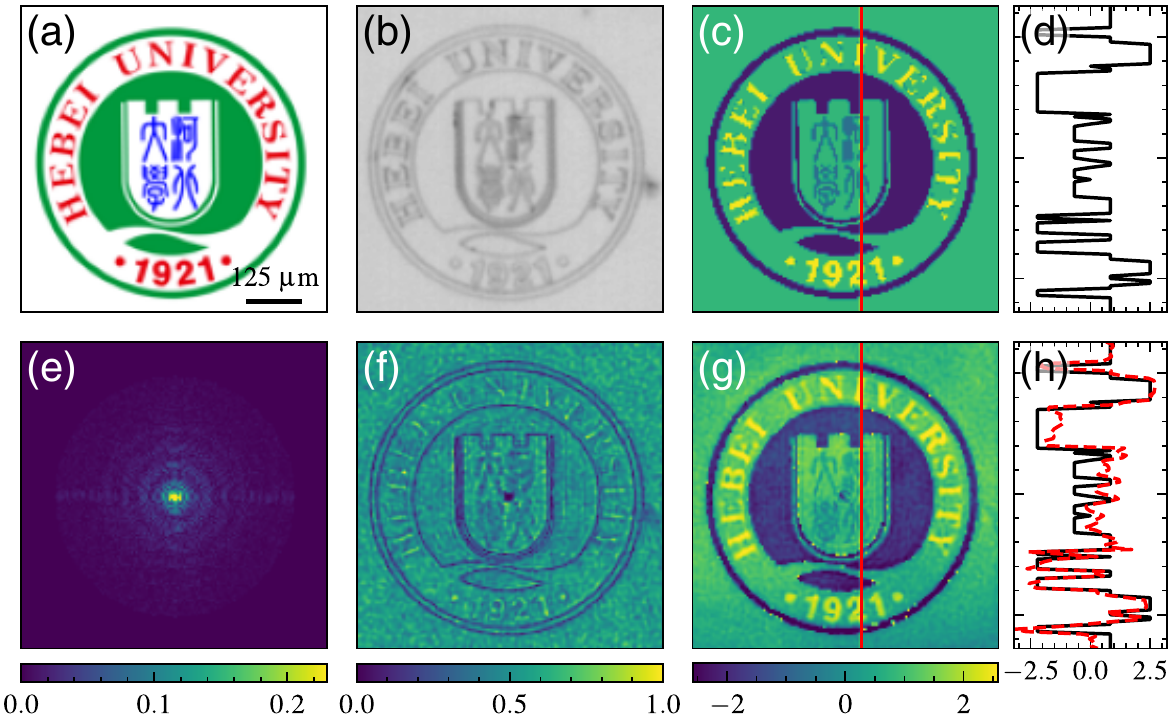}
\caption{\label{fig_logo}The glass plate etched with the logo of Hebei
  University and related experimental results.
  (a) Geometry diagram of the etched logo.
  (b) The macro photo of the etched object.
  (c) The theoretical phase distribution of the etched object.
  (d) The phase profile along the highlighted line shown in (c).
  (e)--(h) are the magnitude of the measured spectrum, the reconstructed
  amplitude and phase images, and the phase profile at the sampling ratio of
  $ 47.9 $\%, respectively. See Data File 2 for underlying values.}
\end{figure}

In the second experiment, we challenge our CFSI with a detailed phase object,
which is a glass plate etched with the logo of Hebei University.
As shown in Fig. \ref{fig_logo}(a), the diameter of the logo is
$ 600 $ {\textmu}m, and the etching depths in the red, green and blue regions
are 372 nm, 715 nm, and 1051 nm, respectively.
Fig. \ref{fig_logo}(b) shows the macro photo of this etched object
captured by a digital camera.
Fig. \ref{fig_logo}(c) shows the theoretical phase distribution of this
etched object for the light of wavelength $632.8$ nm,
and the phase profile along the highlighted line is shown in
Fig. \ref{fig_logo}(d).

In this case, to preserve as much details as possible,
we set the sampling ratio to $47.9$\%.
The imaging acquisition time is about $257$ minutes.
Fig. \ref{fig_logo}(e) shows the magnitude of the measured spectrum.
The reconstructed amplitude and phase images are shown in
Figs. \ref{fig_logo} (f) and (g), respectively,
and the phase profile along the highlighted line is shown in
Fig. \ref{fig_logo}(h).
It can be seen that the reconstructed amplitude image shown in
Fig. \ref{fig_logo}(f) matches very well with the macro photo given in
Fig. \ref{fig_logo}(b).
As seen in Figs. \ref{fig_logo}(g) and \ref{fig_logo}(h) the reconstructed
phase image using this CFSI are clear and accurate.
However, there is a loss of some details in the reconstructed images,
for example, the Chinese characters in the central region of the logo.
This is due to insufficient spatial resolution in the experiment.

\begin{figure}[tbp]
\centering
\includegraphics[width=\columnwidth]{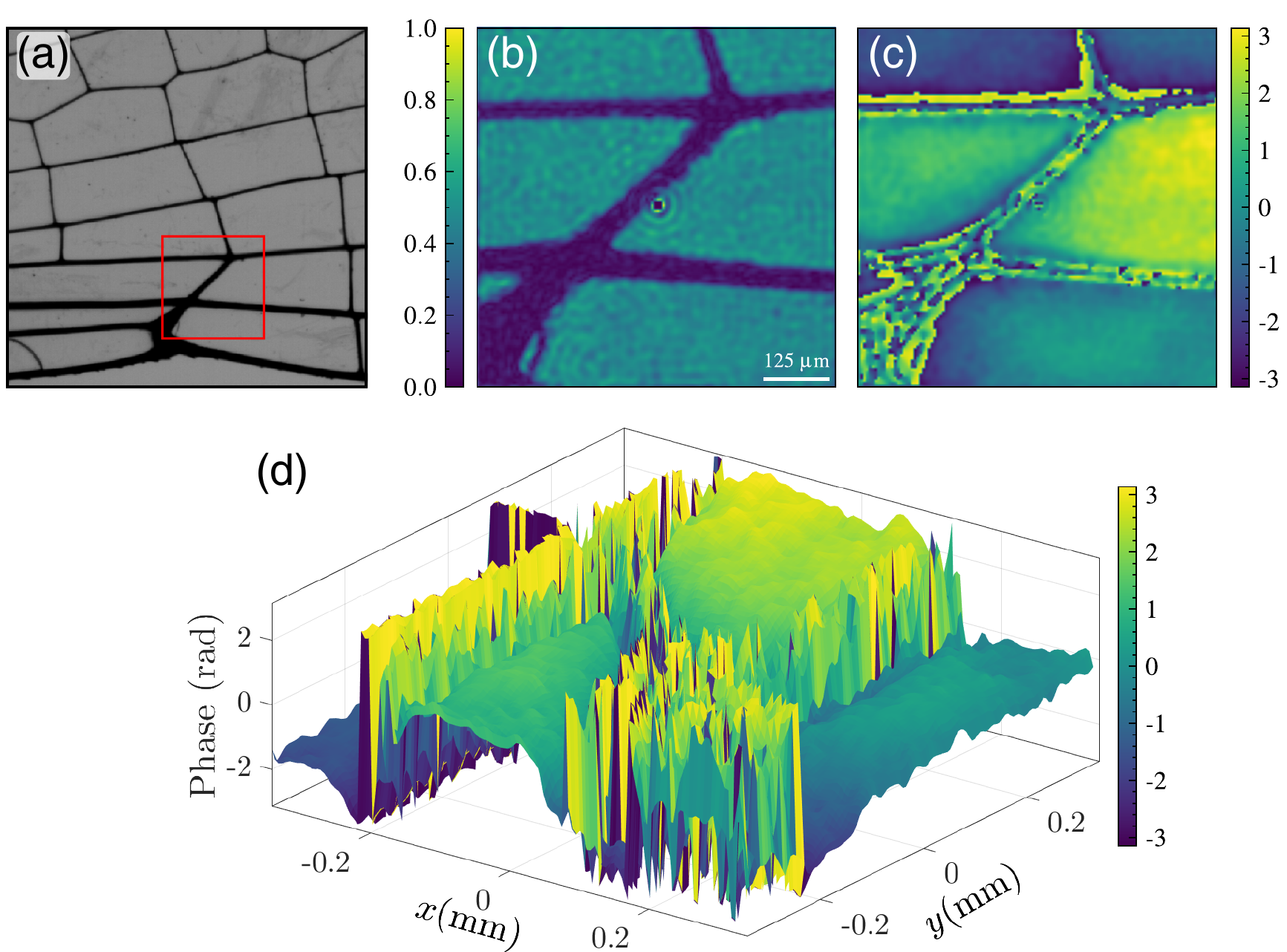}
\caption{\label{fig_wing}The damselfly wing and related experimental results.
  (a) A photo of the wing. The red box indicates the imaging region.
  (b) and (c) are the reconstructed amplitude and phase images, respectively,
  at the sampling ratio of $ 17.2 $\%.
  (d) The corresponding three-dimensional surface of the phase image (c).
See Data File 3 for underlying values.}
\end{figure}

In the third experiment, we use a damselfly wing as the target object,
to test the imaging capability of this CFSI method for natural biological
tissues.
We use this CFSI method to image the wing in the red box region,
as shown in Fig. \ref{fig_wing}(a),
and the reconstructed amplitude and phase images are shown in
Figs. \ref{fig_wing}(b) and \ref{fig_wing}(c), respectively.
The corresponding three-dimensional surface of the phase image is shown in
Fig. \ref{fig_wing}(d).
With the phase distribution, we can obtain more information about the wing.
Due to less details, here we set the sampling ratio to $17.2$\%.
In this case, the image acquisition time is about $34$ minutes.

\begin{figure}[htbp]
\centering
\includegraphics[width=\columnwidth]{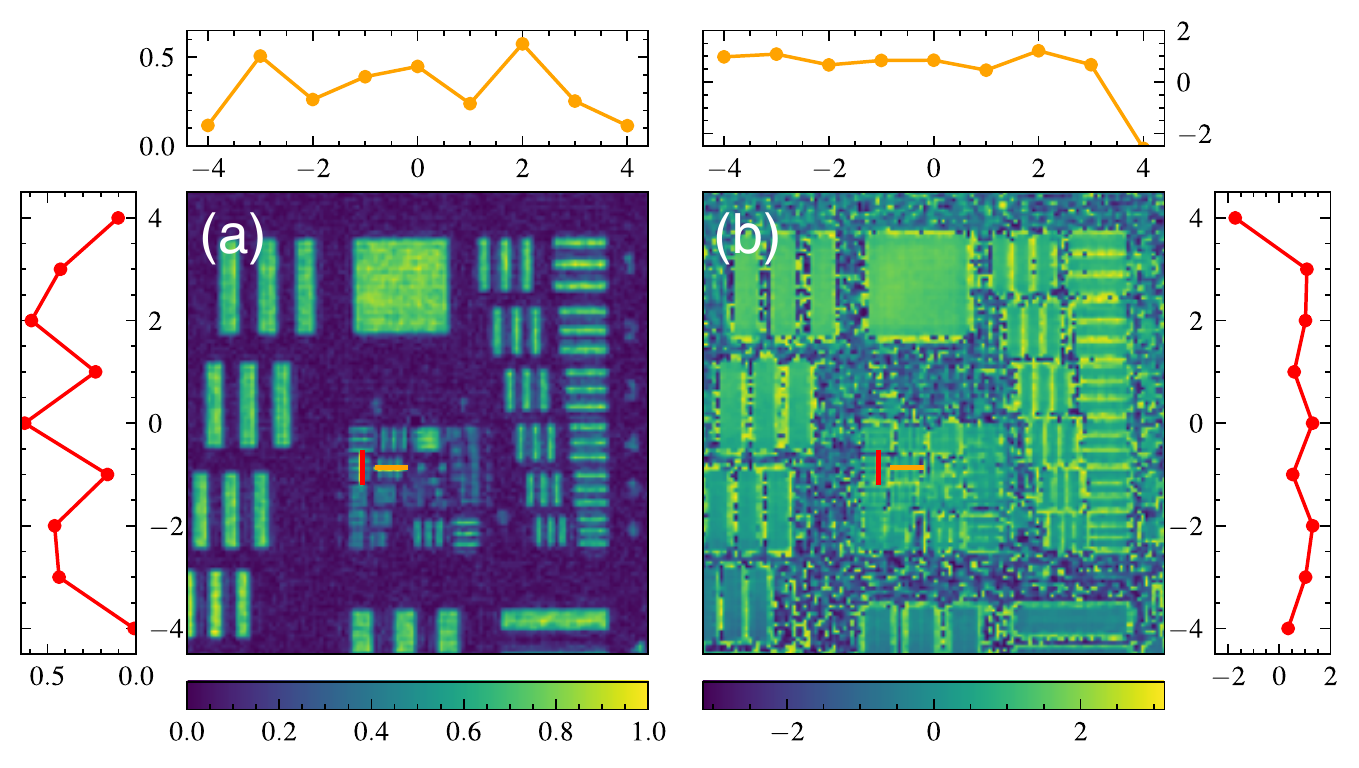}
\caption{\label{fig_usaf1951}Experimental results of the USAF-1951 resolution
  test chart.
  (a) and (b) are the reconstructed amplitude and phase images, respectively
  at the sampling ratio of $ 62.2 $\%.
  The 4 subplots show the amplitude and phase profiles along the highlighted
  lines across the horizontal and vertical bars of group 6 element 3,
  respectively. See Data File 4 for underlying values.}
\end{figure}

Due to vibration, lens aberrations, etc. the achievable spatial resolution
is larger than the theoretical value of $ 5.33 $ {\textmu}m in practice.
In the fourth experiment, we use the negative USAF-1951 resolution test chart
as the target object to quantify the achievable spatial resolution
in the experiments.
The experimental results are shown in Fig. \ref{fig_usaf1951},
where (a) and (b) show the amplitude image and phase image, respectively.
The amplitude profiles along the highlighted lines across the horizontal and
vertical bars of group 6 element 3 are shown in the left and top subplots,
respectively, where the axes along the image represent the pixel positions
and each dot represents the amplitude value of one pixel.
The corresponding phase profiles are also given in Fig. \ref{fig_usaf1951}(b).
As shown in Fig. \ref{fig_usaf1951}(a), group 6 element 3 can be clearly
resolved.
This means that the achievable spatial resolution is about $ 6.2 $ {\textmu}m.
However, since the resolution test chart is a amplitude-only object,
the reconstructed phase in the opaque region is severely
contaminated by noise.
Therefore, the spatial resolution of the phase image does not look good enough.
In this case, we set the sampling ratio to $62.2$\%.
The imaging acquisition time is about $ 334 $ minutes.

\section{CONCLUSION}
In summary, we propose and experimentally demonstrate a novel
complex-amplitude imaging method.
By utilizing a phase-only SLM to generate coherent Fourier basis patterns
as illumination, we can use a single-pixel detector to acquire the
complex-valued Fourier spectrum of an object.
In combination with undersampling technique and noise suppression,
we can recover a high quality complex-amplitude image of the object.
In experiments, we implement this CFSI method with two etched glass objects,
a damselfly wing, and a resolution test chart.
The reconstructed complex-amplitude images have clear amplitude,
accurate phase and spatial resolution of up to $6.2$ {\textmu}m.
In addition, due to the use of common-path interference, the experimental
configuration of this CFSI is compact and stable,
which is readily integrated into commercial microscopes for
quantitative phase microscopy.
Thus, this complex-amplitude imaging method might find broad applications
in optical metrology and biomedical science,
especially for the cases with a light of invisible wavelength
or extremely low intensity.

\begin{acknowledgments}
The authors are grateful to Huan Zhou for her helpful discussions and
suggestions.
This work was supported by
the Natural Science Foundation of Hebei province (F2019201446),
the Multi-Year Research Grant of University of Macau (MYRG2020-00082-IAPME),
the Science and Technology Development Fund from Macau SAR (FDCT)
(0062/2020/AMJ),
the Advanced Talents Incubation Program of the Hebei University (8012605),
and the National Natural Science Foundation of China (NSFC)
(11204062, 61774053, 11674273).
\end{acknowledgments}

\section*{DATA AVAILABILITY}
Data underlying the results presented in this paper are available in
Dataset (Ref. \cite{dataset}).

\bibliography{Reference.bib}

\end{document}